# Buck You: Designing Easy-to-Onboard Blockchain Applications with Zero-Knowledge Login and Sponsored Transactions on Sui

Eason Chen
Carnegie Mellon University

Zimo Xiao
Carnegie Mellon University

Justa Liang
LiquidLogic Tech

Damien Chen
LiquidLogic Tech

Pierce Hung
LiquidLogic Tech

Kostas Kryptos Chalkias
Mysten Labs

Traditional Blockchain Applications
for New Users

Our Application: Zero Knowledge Login with
Sponsored Transaction on Sui

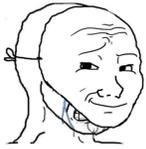 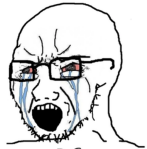 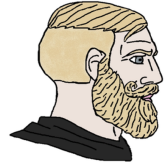 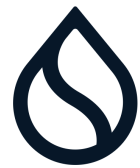

I finally set up my wallet with mnemonic phrases

Noooo, You still need to deposit funds as transaction fee

Let me register with Google by Zero Knowledge Login

Great! Now let's start with Sponsored Transaction

Figure 1: Meme depicts how our application (https://send.buckyou.io) differs from traditional blockchain applications. In our application, new users don't need to undergo the cumbersome wallet creation and gas preparation process.

## ABSTRACT

In this paper, we developed a blockchain application to demonstrate the functionality of Sui's recent innovations: Zero Knowledge Login and Sponsored Transactions. Zero Knowledge Login allows users to create and access their blockchain wallets just with their OAuth accounts (e.g., Google, Facebook, Twitch), while Sponsored Transactions eliminate the need for users to prepare transaction fees, as they can delegate fees to sponsors' accounts. Additionally, thanks to Sui's Storage Rebate feature, sponsors in Sponsored Transactions can profit from the sponsorship, achieving a win-win and sustainable service model. Zero Knowledge Login and Sponsored Transactions are pivotal in overcoming key challenges novice blockchain users face, particularly in managing private keys and depositing initial transaction fees. By addressing these blockchain user experience challenges, Sui makes blockchain more accessible and engaging for novice users and paves the way for the broader adoption of blockchain applications in everyday life.

## 1 INTRODUCTION

In recent years, blockchain technology has become extremely popular, with the assets placed on blockchains and the number of users continually growing. By 2023, blockchain manages assets worth over $54 billion USD equivalent [4], accompanied by an active user base with more than 3.74 million active addresses [2, 3].

Despite the popularity of blockchain, there is still a lot of room for improvement in the user experience[19]. Especially when it comes to engaging new blockchain users [34], issues like private key management and gas fees make blockchain unfriendly for beginners [23, 34]. For example, to create a Self-custody wallet where users manage private keys by themselves, users need to remember their wallet information, such as mnemonic phrases; if users lose this information, they may lose access to their assets on the blockchain [22, 37]. Moreover, the transaction fee problem frequently confuses new users, as they must deposit some funds from centralized exchanges into their blockchain wallets as an initial gas fee before they can send blockchain transactions[22, 26].

The aforementioned challenges have posed significant obstacles when new users engage with blockchain technology. Even though many studies [19, 22, 23, 34, 34, 37] have highlighted several blockchain user experience challenges in the HCI community, few have focused on addressing these challenges and creating applications that demonstrate potential solutions [23]. Our paper aims to fill this gap by sharing the latest technology innovations at Sui, one of the newly introduced blockchains, with real application use case. These innovations include the integration of Zero Knowledge Login [10, 15] (explained at 3.4) for creating Self-custody blockchain wallets without memorizing wallet information, and the implementation of Sponsored Transactions [6] (explained at 3.3) that enables new users to initiate transactions without pre-depositing funds for transaction fee.

To summarize, our contributions are as follows:

(1) We explained how Zero-Knowledge Login and Sponsored Transactions work on Sui in a sustainable way.
(2) We built an application to demonstrate how users could easily start using Sui with their OAuth account (e.g., Google) by Zero-Knowledge Login and Sponsored Transactions without any gas fee preparation.
(3) We discussed how Zero-Knowledge Login and Sponsored Transactions can revolutionize the future blockchain user experience.

Through our contribution, we aim to enhance the HCI community's understanding on blockchain technology's recent advancements and future potential, sparking further discussions and research in this field.

## 2 BACKGROUND AND RELATED WORK

### 2.1 Blockchain and Gas Fee

Blockchain is a distributed ledger technology (DLT) that securely records transactions across a network with multiple nodes, ensuring transparency, immutability, and data trust [31, 38]. A traditional blockchain comprises a series of 'blocks' linked together like a chain. Each block contains a record of transactions and a reference (known as a hash) to the previous block. This reference is crucial as it maintains the integrity of the blockchain, ensuring that each transaction in the block is legit across the network.

Smart contracts are code-based logic that operates on the blockchain. They can perform predetermined actions when certain conditions are fulfilled. Smart contracts are used to automate a wide range of processes and increase trust by removing the risk of human intervention. Smart contracts are an integral part of blockchain networks, and currently, numerous decentralized applications (DApps) have been created on different blockchains with smart contracts.

Every call to a blockchain function, such as writing data or executing smart contracts, requires costs known as transaction fees [31]. This practice is indispensable due to the decentralized nature of the blockchain, where computations and data storage occur across multiple nodes. Additionally, to ensure consistency, transactions must be executed sequentially and packaged into blocks, which limits the computational capacity of the platform. Transaction fees serve as a means to prevent resource misuse and maintain cost equilibrium. In practice, users have to increase the transaction fees to prioritize the confirmation of their transactions when the network is heavily loaded. As a result, calculating transaction fees has become a significant concern for blockchain users [21, 37].

In addition, users must have assets in their wallets for transaction fees to use blockchain. This typically involves purchasing cryptocurrencies on a centralized exchange and then depositing the assets into a blockchain wallet before they can start interacting with DApps on the blockchain. This cumbersome process often discourages potential new blockchain users [34, 37].

Thankfully, several modern, groundbreaking DLT designs, such as the Sui Network, have arisen to diminish the computational and storage demands on the network. These advancements enhance computational efficiency, consequently addressing transaction fee concerns (explained at 3.1). Additionally, Sui offers a sponsored transaction option that enables new users to interact with smart contracts without the preparation of transaction fees (explained at 3.3).

### 2.2 User Experience Challenges in Key Management

In blockchain technology, transaction authentication is achieved independently of central entities like banks through asymmetric cryptography [31], which utilizes a pair of mathematically linked keys: a private key and a public key. The private key, akin to a cryptographic password, is used to access and authorize transactions from the corresponding wallet. The public key, shared openly, serves as an address for receiving crypto-assets. The Private Key can be used to generate a signature, which can be used to derive the Public Key, but it is impossible to reverse engineer the Private Key from the signature nor the Public Key. This cryptographic process, verified through algorithms [12, 25] such as RSA [33], ECDSA [27], DSA [32], and EdDSA [28], ensures that each transaction is legitimate and that the assets involved are controlled by their rightful owner.

In practical use, when setting up their blockchain wallet, users generated a series of "mnemonic phrases" based on the BIP-39 protocol [1] to derive their private keys, making it easier for them to remember. However, copying these mnemonic phrases remains a cumbersome task [13, 34, 37]. Current blockchain wallets typically require users to record mnemonic phrases during setup, which consumes approximately three minutes. Additionally, if the mnemonic is lost, it will result in a loss of access to the wallet.

Apart from this, while the private-public key authentication approach ensures the security of asset access and grants users full control over their assets, it also imposes greater responsibility and challenges on users in managing their keys. These challenges include: confusion distinguishing between public and private keys [20, 21, 34, 37], a lack of confidence in accurately entering the correct key [21, 34], incidents of inadvertently sharing the private key which should remain confidential [13, 20], failures in backing up the private key or passphrases for key recovery [20, 30], and struggles with the lengthy and less human-memorable nature of the key's composition [13, 34, 37]. These challenges can hinder blockchain adoption among non-technical and novice users who might find the process less intuitive and more cumbersome [34, 37].

Lastly, given that the private key is the sole method of authentication and considering the often limited experience and knowledge of novice users in key management, this user experience paradigm creates opportunities for accidental yet consequential self-induced errors, such as misplacing or forgetting their keys [13, 22, 29]. Such mistakes carry the significant risk of irreversible fund loss, underlining a critical need for more user-friendly and error-tolerant approaches in blockchain interface design.

Notably, the recently introduced Zero Knowledge Login by Sui (explained at 3.4) is meant to be the perfect solution to the aforementioned problems.

## 3 RELATED WORKS ON SUI

Mainnet launched in May 2023. Sui is a decentralized and permissionless smart contract platform that focuses on efficient asset management with low latency [36]. Derived from Meta's Diem (formerly known as Facebook's Libra), Sui employs the Move as its smart contract programming language [17] to define and handle assets owned by addresses. It allows for custom rules regarding asset creation, transfer, and mutation.

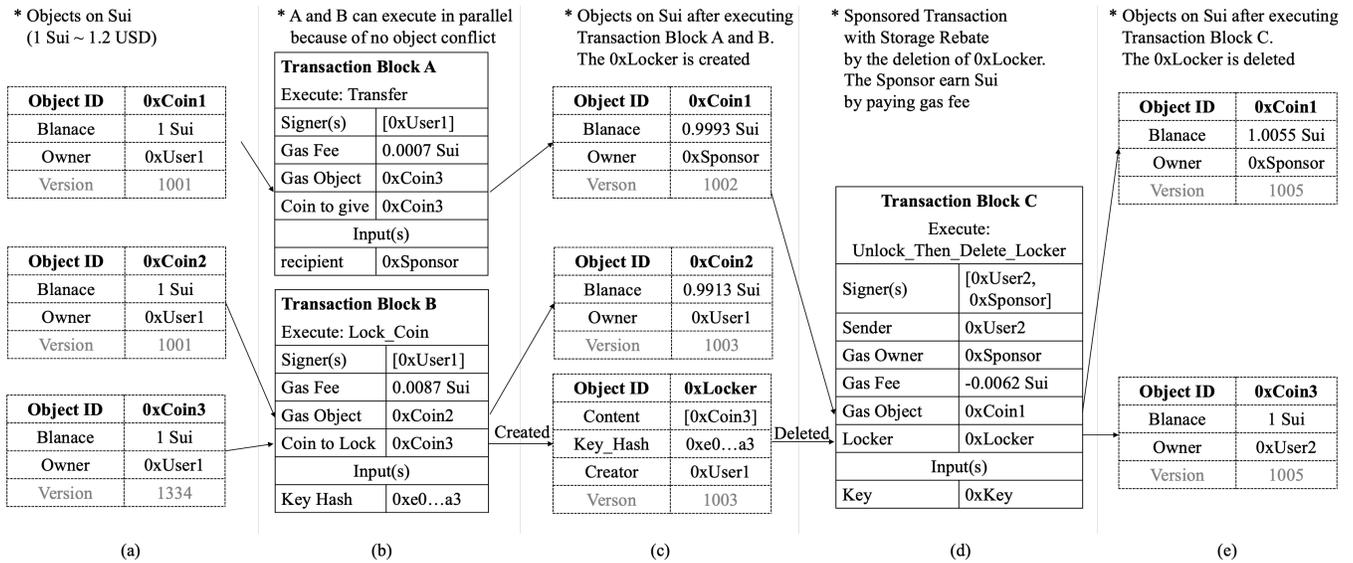

Figure 2: The figure illustrates how object-based transactions work on Sui with the example of our application. In (a), it depicts the initial state of objects on Sui where three Coins belong to 0xUser1. In (b), 0xUser1 initiates two transactions: in Transaction Block A, transferring 1 Sui (0xCoin1) to 0xSponsor, and in Transaction Block B, locking another 1 Sui (0xCoin3) into a smart contract. It's important to note that the transactions can be processed in parallel because object inputs are independent. (c) shows the blockchain state after the transactions are executed, highlighting the creation of a new object, 0xLocker, managed by smart contract. (d) Next, in the transaction, 0xUser2 initiates a transaction with the right key of the locker and requests 0xSponsor to sign the transaction to sponsor the transaction fees. (e) As a result, 0xUser2 successfully interacts with the smart contract and withdraws the locked assets without preparing any funds to pay the transaction fees. Additionally, 0xSponsor's assets increase after sponsorship due to the Storage Rebate.

### 3.1 Sui's Object-Based and Directed Acyclic Graph Design

Unlike Ethereum or Solana's account-based system, where assets are represented as numerical variables within addresses or smart contracts, Sui uses an object-based approach to manage assets, similar to Bitcoin's Unspent Transaction Output (UTXO) structure [31]. This enables digital assets to be fragmented, combined, and transferred to different addresses.

Moreover, what sets Sui apart from traditional blockchains is its use of a Directed Acyclic Graph (DAG) model to manage objects and record transactions [8]. As illustrated in Figure 2, each transaction block on Sui contains transactions with inputs from various objects on the Sui Network. These transactions then create or modify objects. Thanks to the DAG and object-based design, Sui allows transactions involving unrelated objects without a specific sequence [8], as shown in Figure 2 (b). This maximizes Sui's computational efficiency and scalability. Consequently, despite Sui experiencing a rapid increase in transaction volume from thousands to tens of millions in a short period, Sui's transaction fees have remained relatively stable and considerably low [18].

### 3.2 Storage Rebate on Sui

In addition, Sui has made optimizations in data storage. Sui allows the deletion of data on the network to free up space and obtain a Storage Rebate fee [7]. This ultimately reduces the overall long-term cost of storing a substantial amount of data on Sui, leaving only the computational and necessary storage fees, as extra storage costs can be rebated by clearing data storage after operations are completed. For example, in Figure 2 (b), Transaction Block B created a Locker Object on Sui with 0.0087 Sui as the transaction gas fee. Then, in Figure 2 (d), the subsequent Transaction Block C deleted the Locker Object and thus free spaces and got Storage Rebate, resulting in a negative gas fee of -0.0062.

Notably, even though objects are deleted, verification and reproduction remain possible since inputs are stored within the Transaction Block and retained in a database on the Sui nodes without consuming hot-memory resources.

### 3.3 Sponsored Transaction on Sui

In addition, in Sui's design, each Transaction Block allows object inputs from multiple owners as long as these owners provide valid signature authentication for the transaction [5]. This empowered the potential of sponsored transactions that let another address to pay the transaction fee. As illustrated in Figure 2 (d), when 0xUser2, a newly onboarded blockchain user, initiates a transaction, 0xSponsor provides the Gas Coin Object (0xCoin1) as a transaction

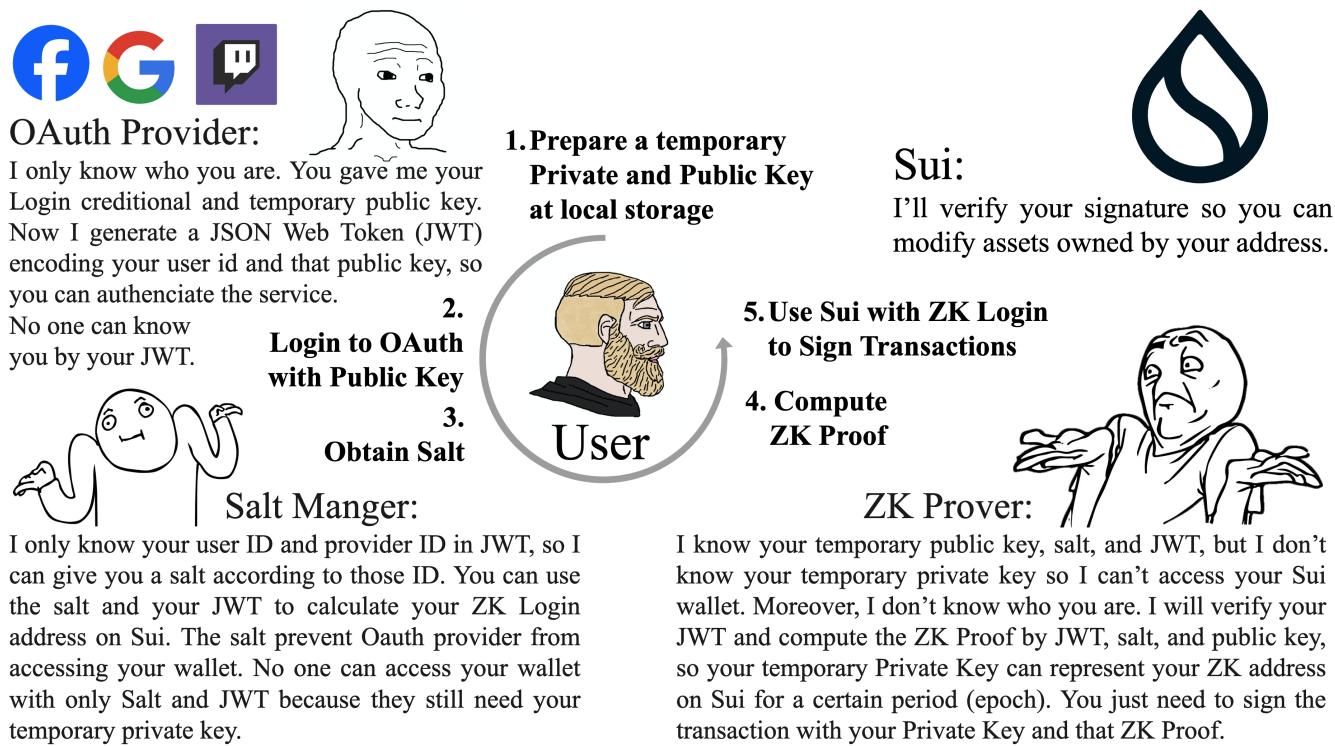

**OAuth Provider:**
I only know who you are. You gave me your Login creditional and temporary public key. Now I generate a JSON Web Token (JWT) encoding your user id and that public key, so you can authenciate the service.
No one can know you by your JWT.

**1. Prepare a temporary Private and Public Key at local storage**

**Sui:**
I'll verify your signature so you can modify assets owned by your address.

**2. Login to OAuth with Public Key**

**5. Use Sui with ZK Login to Sign Transactions**

**3. Obtain Salt**

**4. Compute ZK Proof**

**User**

**Salt Manger:**
I only know your user ID and provider ID in JWT, so I can give you a salt according to those ID. You can use the salt and your JWT to calculate your ZK Login address on Sui. The salt prevent Oauth provider from accessing your wallet. No one can access your wallet with only Salt and JWT because they still need your temporary private key.

**ZK Prover:**
I know your temporary public key, salt, and JWT, but I don't know your temporary private key so I can't access your Sui wallet. Moreover, I don't know who you are. I will verify your JWT and compute the ZK Proof by JWT, salt, and public key, so your temporary Private Key can represent your ZK address on Sui for a certain period (epoch). You just need to sign the transaction with your Private Key and that ZK Proof.

**Figure 3: Figure depicted how Zero Knowledge Login works on Sui and how different roles participate with limited user information.**

gas fee and signs the transaction, thereby allowing the 0xSponsor to cover the gas fee while 0xUser2 send the transaction.

In the past, when designing wallets, attempts regarding sponsored transactions often stalled due to the expensive and unpredictable transaction fees. However, with Sui combining Sponsored Transactions with the previously mentioned Storage Rebate feature, as shown in Figure 2 (d), sponsors actually earn additional Sui tokens by sponsoring transactions, achieving sustainable and win-win results, where new users don't need to prepare the gas fee, while sponsor get profit by sponsorship.

## 3.4 Zero Knowledge Login on Sui

To access a wallet on Sui, in addition to the traditional Private-Public Key authentication, it also supports Zero Knowledge Login (ZK Login) [11, 15], which empowered users to directly use OAuth accounts such as Google, Facebook, or Twitch to connect to their own wallets (Figure 3).

Zero Knowledge (ZK) here originates from the cryptographic concept of Zero Knowledge Proof (ZKP). ZKP is a method whereby one party (the prover) can prove to another party (the verifier) that a given statement is true, without conveying any information beyond the mere fact of the statement's truth.

Figure 3 illustrated the flow of Zero Knowledge Login on Sui and how different roles participate with limited user information. Each step is explained as follows:

(1) **Prepare Temporary Key**: The user prepares a temporary Private and Public Key in local storage (usually in the browser).

(2) **OAuth Login**: The user uses this Public Key to initiate an OAuth Login with the OAuth Provider. After verifying the user's identity through account authentication, OAuth generated a JSON Web Token (JWT) for the user. This JWT contains the user's account information and can be verified using the corresponding Public Key. According to the OAuth management protocol, without user authorization, the OAuth provider cannot generate a JWT on its own, and others cannot access the user's personal information through the JWT by default[1].

(3) **Obtain Salt**: The user provides the JWT to the Salt Manager to obtain their own Salt. Then, they use the Salt and the personal information from the JWT to calculate their address. This Salt helps prevent the OAuth provider from accessing the user's account when a hacker compromises the OAuth provider.

(4) **Calculate Zero Knowledge Proof (ZK Proof)**: ZK Proof is a native signature verification method on Sui, which can be generated by the user's JWT, Salt, and Temporary Public Key and is only valid before the JWT expires. While users can generate ZK Proof on their own, typically, due to performance considerations, ZK Proof is generated by a backend server equipped with sufficient computational resources (referred to as the ZK Prover). Although the ZK Prover knows the JWT, Salt, and Public Key, it can't access the user's wallet because it doesn't have the Private Key.

---

[1]Users may further configure with the OAuth provider if they agree to share personal information with others who have their JWT.

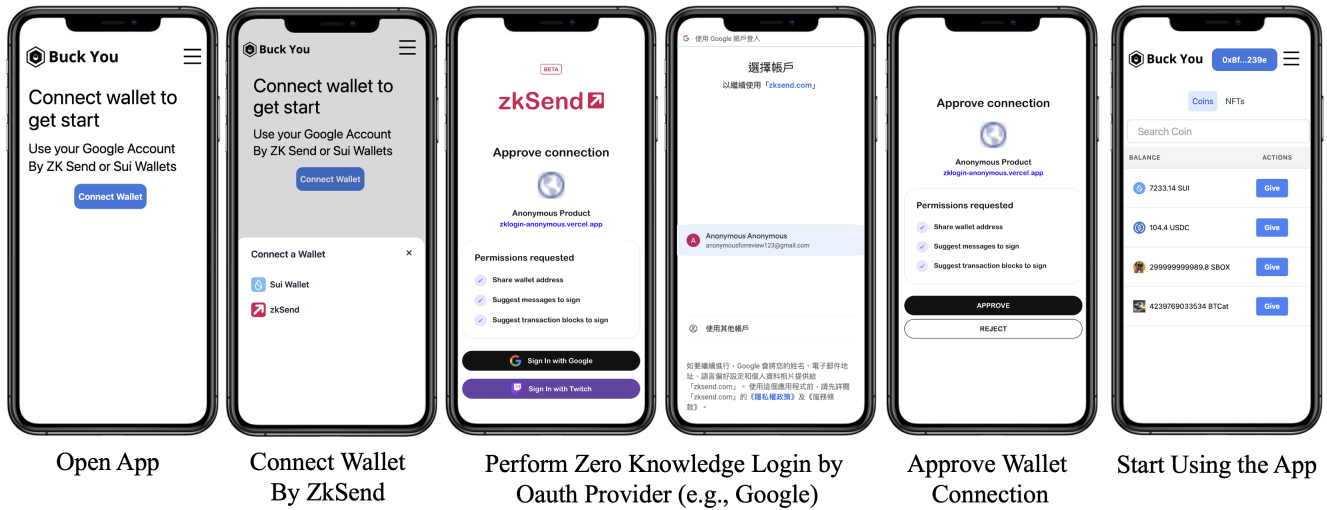

| Open App | Connect Wallet By ZkSend | Perform Zero Knowledge Login by Oauth Provider (e.g., Google) | Approve Wallet Connection | Start Using the App |

**Figure 4: Screenshot of how users connect their wallet by ZK Login at our application. They first click the "Connect Wallet" button, then select "zkSend", and login with their OAuth provider (e.g., Google). Finally, they can connect their ZK Login wallet to our application and use it like a normal wallet.**

(5) **Sign Transaction on Sui**: When the user signs a transaction with the Temporary Private Key and ZK Proof, it can be verified as a legitimate signature on Sui. As a result, users can modify assets owned by their address with the signature from ZK Login.

To sum up, ZK Login offers the advantage of ensuring user wallet autonomy by their OAuth account while preventing unauthorized access to the user's wallet, thus achieving privacy and security assurance. For more information about ZK Login, please refer to Sui's official documentation [11] and its paper [15].

To try ZK Login, please visit our application at https://send.buckyou.io and use "Connect by zkSend". Figure 4 illustrates our application's user interface for performing a ZK Login. It takes less than twenty seconds for a new user to perform ZK Login and access their own wallet, significantly quicker than traditional methods in which users need to set up a private key wallet and handle mnemonic phrases. This time is comparable to the time users spend to register a web application by OAuth, such as registering Overleaf by Google.

## 4 APPLICATION DESIGN

Our application (https://send.buckyou.io) integrates ZK Login with Sponsor Transaction for users to give away assets on Sui to other users. ZK Login allows users to connect their wallets with their OAuth account, such as Google. Subsequently, Sponsor Transaction enables users to interact with smart contracts immediately with their new account without depositing a transaction fee.

*4.0.1 Current Progress.* Our application has been launched on https://send.buckyou.io. Please play around at the website.

### 4.1 User Flow

The screenshot of User Flow is at Figure 5. The purpose of our application is to enable blockchain users to quickly onboard new blockchain users. In the application, the sender can select the assets they want to send to the receiver. These assets are then locked in a smart contract using the hash of a Key. The sender then displays a QR Code for the receiver to scan, which contains information about the Key. Once the receiver scans it, they can use this key to call the contract to unlock the assets inside through a sponsored transaction and receive the sender's gift. This enables the recipient to receive the sender's gift (e.g., NFTs and Coins) on Sui, even if the recipient does not yet possess a Sui wallet.

### 4.2 Smart Contract Design

The contract call process is depicted in Figure 2. We use the Hash function SHA3_256 to ensure the sender does not reveal the Key when depositing assets. Hash is a one-way function in which we can generate a Key Hash from the Key, but retrieving the Key from the Key Hash is nearly impossible. Additionally, we employ Sui's Object Bag to store the items placed inside it. When the Receiver retrieves items from the Object Bag, the Dynamic Object Fields within the Object Bag are deleted, achieving the effect of Storage Rebate. Please check attachment for smart contract source code.

## 5 DISCUSSIONS, LIMITATIONS, AND FUTURE WORKS

In our paper, we build an application that demonstrates how Zero Knowledge Login (ZK Login) and Sponsored Transaction on Sui can eliminate the hassle of setting up a new blockchain wallet and depositing funds for new users sustainably. These two features

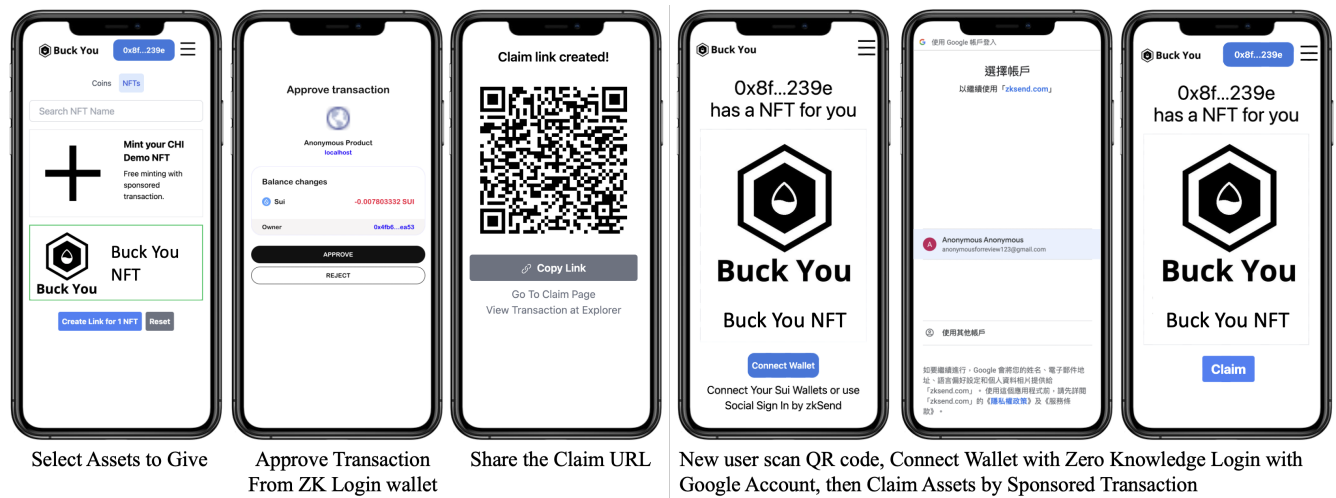

| Select Assets to Give | Approve Transaction From ZK Login wallet | Share the Claim URL | New user scan QR code, Connect Wallet with Zero Knowledge Login with Google Account, then Claim Assets by Sponsored Transaction |

**Figure 5: Screenshot of user flow of our application, which enables users to send NFT to others by QR Code, then empowered new users to perform Zero Knowledge Login by their Google account and claim NFT with sponsored transaction**

overcome the traditional hassle for new users in which they need to handle mnemonic phrases to blockchain wallets and go through intricate steps to pre-deposit fees to interact with the blockchain.

ZK Login and Sponsored Transaction allow blockchain's transparent, immutable, and automated nature to be integrated into general users' everyday lives. By integrating ZK Login and Sponsored Transaction, future blockchain applications can empower users to quickly connect to the blockchain and utilize smart contracts for activities like donation [16], management[24], voting[14], and gaming. For example, users can swiftly register a blockchain wallet for anonymous voting while still being authenticated as a community member through zero-knowledge proofs. We look forward to seeing these technologies applied and researched within the HCI community to create more user-friendly blockchain applications in the future.

Lastly, it is important to note that the current version of our application still has some limitations that need to be improved, especially the address binding limitations. This is because the user's OAuth account generates the address. If users lose access to their OAuth account, they will lose access to their ZK Login wallet on Sui. To address this limitation, Sui is currently developing an account abstraction approach [35] that a virtual wallet managed by smart contract [9] can be accessed by multiple signers, empowering users to bind many OAuth accounts or even private keys and mnemonic phrases to their ZK Login Sui wallet.

## 6 CONCLUSION

In this paper, We developed an application to demonstrate the function of Sui's recent innovations: Zero Knowledge Login and Sponsored Transactions. These innovations are pivotal in overcoming key challenges faced by new blockchain users, particularly in managing private keys and depositing initial transaction fees. By addressing these challenges, Sui paves the way for a broader blockchain application in daily life, making it more accessible and engaging for users.